\documentclass[showpacs,preprintnumbers,amsmath,amssymb,a4paper]{revtex4}
\usepackage{graphicx}% Include figure files
\usepackage{dcolumn}% Align table columns on decimal point
\usepackage{bm}% bold math
\usepackage{epsfig}
\usepackage{colordvi}

\newcounter{fig}

\newcounter{One}
\newcounter{Two}
\newcounter{Three}
\def\beq{\begin{equation}}
\def\eeq{\end{equation}}
\newcommand{\gesim}{\lower.7ex\hbox{$\;\stackrel{\textstyle>}{\sim}\;$}}
\newcommand{\lesim}{\lower.7ex\hbox{$\;\stackrel{\textstyle<}{\sim}\;$}}

\begin{document}

\title{A quantum mechanical description of the experiment on the observation
of gravitationally bound states}

\author{A. Westphal}
\affiliation{Physikalisches Institut der Universit\"{a}t
Heidelberg\\ Philosophenweg 12\\ 69120 Heidelberg, Germany}
\altaffiliation[Corresponding author. ]{Email address:
westphal@sissa.it\\Present address: ISAS-SISSA and INFN, Via
Beirut 2-4, I-34014 Trieste, Italy}

\author{H. Abele}
\affiliation{%
Physikalisches Institut der Universit\"{a}t Heidelberg\\
Philosophenweg 12\\
69120 Heidelberg, Germany
}%

\author{S. Bae\ss ler}%
\affiliation{%
Institut f\"ur Physik, Universit\"{a}t Mainz\\
Staudinger Weg 7\\
55128 Mainz, Germany
}%

\author{V.V. Nesvizhevsky and A.K. Petukhov}%
\affiliation{%
Institut Laue langevin, 6 rue Jules Horowitz, 38042, Grenoble, France
}%

\author{K.V. Protasov}%
\affiliation{%
LPSC, IN2P3-CNRS, UJFG, 53, Avenue des Martyrs, 38026 Grenoble}%

\author{A.Yu. Voronin}%
\affiliation{%
P.N. Lebedev Physical Institute, 53 Leninsky Prospekt, 119991, Moscow, Russia\\ }%

\date{February 10, 2006}

\begin{abstract}
Quantum states in the Earth's gravitational field were observed,
when ultra-cold neutrons fall under gravity. The experimental
results can be described by the quantum mechanical scattering
model as it is presented here. We also discuss other geometries of
the experimental setup which correspond to the absence or the
reversion of gravity. Since our quantum mechanical model
describes, particularly, the experimentally realized situation of
reversed gravity quantitatively, we can practically rule out
alternative explanations of the quantum states in terms of pure
confinement effects.
\end{abstract}


\maketitle

\section{Introduction}
A gravitationally bound quantum system has been realized
experimentally with ultra-cold neutrons falling under gravity and
reflecting off a "neutron
mirror"~\cite{Nesvizhevsky1,Nesvizhevsky2}. UCN are neutrons
which, in contrast to faster neutrons, are reflected at all angles
of incidence. For such UCN, flat surfaces thus act as mirrors.
Using an efficient neutron absorber for the removal of higher
unwanted states, only neutrons in selected energy states are
taken. This idea of observing quantum effects occurring when
ultra-cold neutrons are stored on a plane matter surface was first
discussed by V.I. Lushikov and I.A. Frank~\cite{Lushikov} with the
first concrete experimental realization in~\cite{Nes2000}. An
experiment in some aspects similar was discussed by H. Wallis et
al.~\cite{Wallis} in the context of trapping atoms in a
gravitational cavity. The toy model of a Schr\"odinger quantum
particle bouncing in a linear gravitational field is known as the
quantum bouncer~\cite{Gibbs,Rosu}. Retroreflectors for atoms have
used the electric dipole force in an evanescent light
wave~\cite{Aminoff,Kasevich} or they are based on the gradient of
the magnetic dipole interaction, which has the advantage of not
requiring a laser~\cite{Roach}.

A unique side-effect of the experiment with neutrons is its
sensitivity to gravity-like forces at length scales below 10
$\mu$m while all electromagnetic effects are extremely
suppressed~\cite{Abele,Prot}. The quantum states probe Newtonian
gravity between $10^{-9}$ and $10^{-5}$ m and the experiment
places limits for gravity-like forces there. In light of recent
theoretical developments in higher-dimensional field
theory~\cite{Arkani1} (see also~\cite{stringADD} for explicit
realizations in string theory), gauge fields could mediate forces
that are $10^{6}$ to $10^{12}$ times stronger than gravity at
submillimeter distances, exactly in the interesting range of this
experiment and might give a signal in an improved setup.

In this article, we provide the details of a quantum mechanical
calculation~\cite{Westphal2} for our experiment, where
gravitational bound quantum states are observed for the first
time. The experiment consists of a reflector for neutrons, called
neutron mirror, an absorber for unwanted neutrons and a neutron
detector. In our previous papers
~\cite{Nesvizhevsky1,Nesvizhevsky2,Nes05} the experiment and a
first treatment of the data were presented. Fundamental limits for
the spatial resolution and a first ansatz to incorporate the
neutron scatterer can be found in~\cite{Nes05}. In another
work~\cite{Voronin} a description of the neutron loss from first
principles was developed where the rough edges of the absorber
surface are treated as a time-dependent variation of a \emph{flat}
absorber position, modeling the neutron loss mechanism as a
process equivalent to the ionization of a particle, initially
confined in a well with an oscillating wall. Within the older and
more simple model we present in this paper we are able to describe
the experimental data with one micro-physical fit parameter which
parameterizes the micro-physics of the neutron scatterer/absorber.
At the moment our model is the only one yet in which both the data
in Fig. 2 and Fig. 4 are described.

\begin{figure}[ht]
\begin{minipage}{18pc}
\includegraphics[width=18pc]{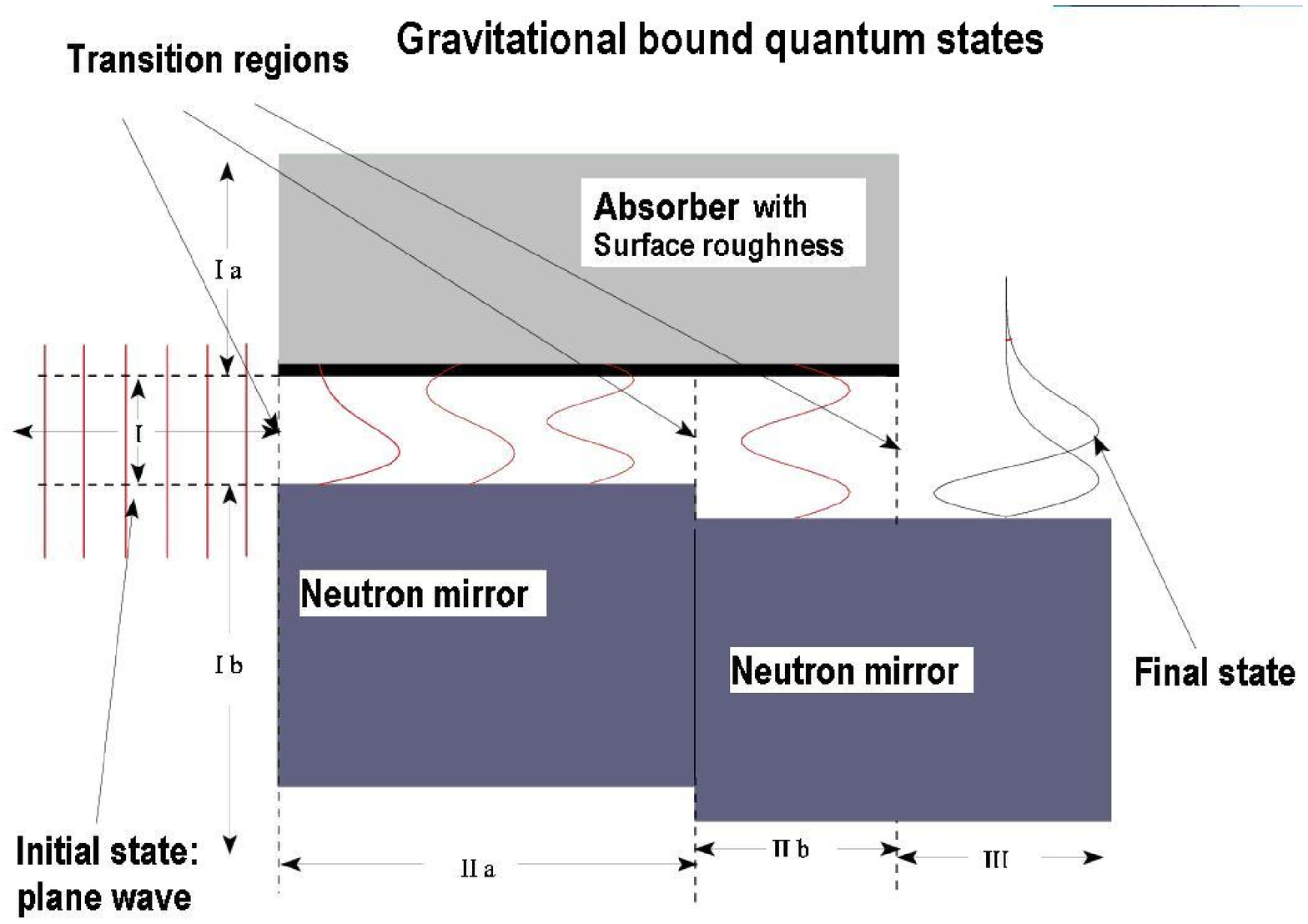}
\caption{\label{Fig.1}Schematic view with mirrors, absorber and quantum
mechanical boundary conditions. In the experiment, one mirror of length 10 cm
or, as an option as shown here, two bottom mirrors of length 6 cm were used. }
\end{minipage}\hspace{1pc}%
\begin{minipage}{18pc}
\includegraphics[width=18pc]{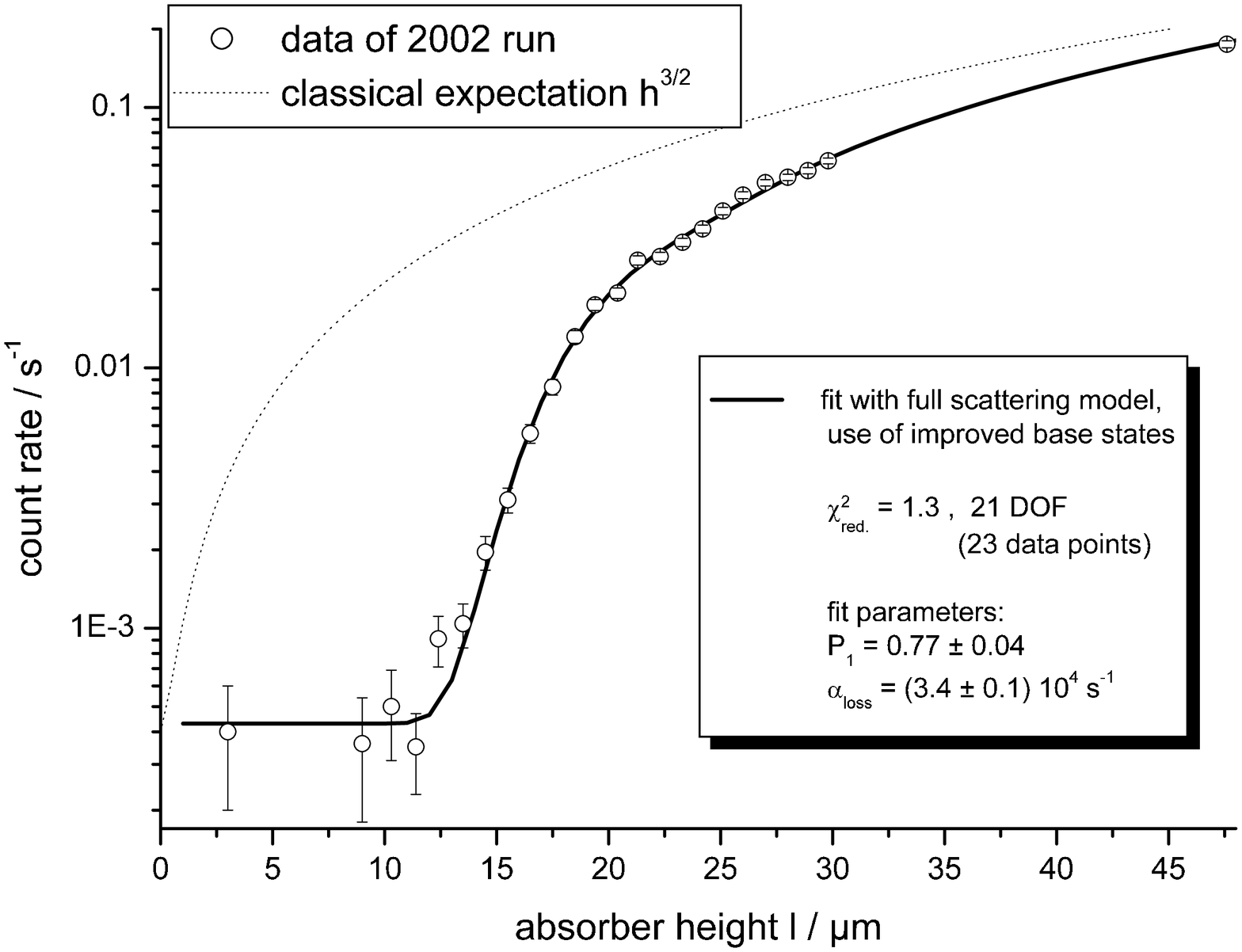}\vspace{-0.3cm}
\caption{\label{Fig.2}\small Circles: Data from the $2^{nd}$ run
2002 with one bottom mirror~\cite{Nes05}. Solid: Transmission
coefficient from the phenomenological scattering model. Dash: The
classical expectation for the neutron transmission coefficient.}
\end{minipage}
\end{figure}

The layout of this paper is the following: After describing the experiment and
the observations, we recall the quantum mechanics of gravitational bound states
on a free mirror, that is without an absorber/scatterer. Then we present an
approach to describe the state selection by deriving a neutron loss rate due to
non-specular scattering from the rough surface of our absorber. This approach
explains the non-classical dependence of the transmission of the
mirror-absorber system as a function of the height $l$ of the absorber above
the mirror, see Fig.~\ref{Fig.1} and Fig.~\ref{Fig.2}. We take into account the
deformation of the bound state wave functions due to the matter bulk of the
state selector, and compare the full prediction with actual data. We also show
that the data are only understood when gravity is present.

\section{Observation of quantum states and Setup}
A description of the experiment at the Institute Laue-Langevin
(ILL) can be found in~\cite{Nesvizhevsky2}. It was installed at
the UCN facility PF2 of the Institute. Here, neutrons have a
velocity of several meters per second. They are then guided to the
experiment via a curved neutron guide with a diameter of 80 mm. At
the entrance of the experiment, a collimation system cuts down to
an adjustable transversal energy $E_{\perp}$ in the pico-eV range.
Either one solid block with dimensions 10 cm $\times$ 10 cm
$\times$ 3 cm or two solid blocks with dimensions 10 cm $\times$ 6
cm $\times$ 3 cm composed of optical glass serve as mirrors for
UCN neutron reflection. To select different states an
absorber/scatterer, a rough mirror coated with an Neutron
absorbing alloy, is placed above the first mirror. We can vary the
height $l$ above the mirror, that is the size of the slit. The
collimation system in front of the state selector is adjusted in
such a way that neutrons on classical trajectories entering the
experiment have to hit the mirror surface at least two times.
After the second mirror we placed a $^{3}$He counter for neutron
detection. Fig. 1 shows a schematic view of our setup. Signatures
of quantum states are observed in the following way: The $^{3}$He
counter measures the total neutron transmission $F$, when neutrons
are traversing the mirror absorber-system as described. The
transmission is measured as a function of the absorber height $l$
and thus as a function of neutron energy since the height acts as
a selector for the energy $E_{\perp}$ of the vertical motion. From
the classical point of view, the transmission $F$ of neutrons is
proportional to the phase space volume allowed by the absorber. It
is governed by a power law $F\sim l^n$ and $n = 3/2$.

The measurements show the following: Above an Absorber/Scatterer height of
about 60 $\mu$m, the measured transmission is in agreement with the classical
expectation but below 50 $\mu$m, a deviation is clearly visible. Below about 15
$\mu$m, no neutrons can pass the slit. In the next section we will find that
the vertical extension of the gravitational bound states increases with the
quantum number. Ideally, we expect a stepwise dependence of $F$ as a function
of $l$. If $l$ is smaller than the spatial width of the lowest quantum state,
then $F$ will be zero. When $l$ is equal to the spatial width of the lowest
quantum state $F$ will increase sharply. A further increase in $l$ should not
increase $F$ as long as $l$ is smaller than the spatial width of the second
quantum state. Then again, $F$ should increase stepwise. At sufficiently high
slit width one approaches the classical dependence. Fig. 2 shows details of the
quantum regime below an absorber height of $l$ = 50 $\mu$m. The transmission
function depends on the horizontal neutron velocity and the absorption
efficiency. It was found, that except for the ground state, the stepwise
increase is mostly washed out.

\section{Quantum mechanical description of gravitationally bound states}

The quantum mechanical treatment of a reflecting neutron mirror, made from
glass, is simple. The glass is described by a Fermi pseudo-potential
($V-i\,W$). This potential is essentially real ($|W|<<|V|$) because of the
small absorption cross section of glass and $V$ = 100 neV is large compared
with the transversal energy $E_{\perp}$ of the neutrons. Therefore, the
potential $V$ is set to infinity at height z = 0. Neutrons which hit the glass
surface undergo specular reflections.

We start with the description of the free states. On a perfect
mirror, no mixing of momentum components take place, which leads
to a decoupled one-dimensional stationary Schr\"odinger equation,
\begin{eqnarray}
\left(-\frac{\hbar^{2}}{2m}\bigtriangleup+V(z)\right)\,\Psi_n=E_ n
\, \Psi_n \label{Eq.01b} , V\left(z\right) =
\left\{\begin{array}{r@{\quad:\quad}l}
mgz  &  x\ge 0\\
\infty & x<0
\end{array}\right. \label{Eq.02}\end{eqnarray} with wave functions
$\Psi_n$ for energies $E_n$ and the potential $V(z)$. $m$ is the
mass of the neutron and $g$ is the acceleration in the earth's
gravitational field. It is convenient to use
\begin{equation}
\zeta=\frac{z}{R}\;\; \hbox{and, above the mirror,} \;\;
V=m\,g\,R\,\zeta.
\end{equation}
Here, $R$ is a scaling factor, defined as
\begin{equation}
R=\left(\frac{\hbar^2}{2m^2g}\right)^{1/3} \label{Eq.03}.
\end{equation}
Solutions $\Psi_{n,g}\left(\vec{r},t\right)$ of Eq.~(\ref{Eq.01b})
are obtained with an Airy function
\begin{eqnarray}
\Psi_{n,g}\left(\vec{r},t\right)&=
\phi\left(x,y\right)\,\psi_{n,g}\left(\zeta\right)\,e^{-\,i\frac{E_n}{\hbar}t}
\nonumber \\
\psi_{n,g}\left(\zeta\right)&=Ai\left(\zeta-\zeta_n\right)\label{Eq.04}
\end{eqnarray}
The displacement $\zeta_n$ of the n-th eigenfunction has to coincide with the
n-th zero of the Airy function (Ai(-$\zeta_n$)=0) to fulfill the boundary
condition $\Psi_n\left(0\right)$=0 at the mirror. The corresponding energies
$E_n$ with $z_n=R\zeta_n$ are
\begin{equation}
E_n=mgz_n\;\;. \label{Eq.06}
\end{equation}
In the WKB-approximation we have to leading order
\begin{equation}
\zeta_n=\left(\frac{3\pi}{2}\left(n-\frac{1}{4}\right)\right)^{2/3}
\label{Eq.06b}
\end{equation}
which coincides with the exact eigenvalues to better than 1 \% even for the
ground state~\cite{Westphal2}. The $z_n$ correspond to the highest point of a
classical neutron trajectory with energy $E_n$. For example, energies of the
lowest levels (n = 1, 2, 3, 4) are 1.44 peV, 2.53~peV, 3.42 peV and 4.21 peV.
The corresponding classical turning points $z_n$ are 13.7 $\mu m$, 24.1 $\mu
m$, 32.5 $\mu m$ and 40.1 $\mu m$.

The aim of this experiment was to populate only some of the lowest allowed
gravitationally bound quantum states. Higher states were removed with the
absorber/scatterer at a certain height $l$.

\section{Phenomenological scattering model of neutron loss}

It will be most convenient to start with a polychromatic neutron beam of
(locally) plane waves entering the system. As is well known, gaussian wave
packets being closer to a particle view of a neutron can easily be decomposed
into a Fourier integral over plane waves. Using all boundary conditions given
by Fig. 1, one arrives at a set of matching conditions~\cite{Westphal2}. The
neutron transmission of the system depends both on the eigenvalues and the
matching conditions. Furthermore, if two bottom mirrors are used and shifted
relative to each other a few $\mu m$ in height (as it was in the 1999 beam
time), there is an additional boundary that would change the population of the
eigenstates.

The insertion of the states yields a system of linear equations for the
matching constants and its solution yields finally the transmission coefficient
of the $n^{th}$ final bound state of region $III$. The initial population of
the bound states of the wave guide system at coupling-out is uniform, if the
vertical velocity distribution of the arriving beam is sufficiently wide and
flat - which is of course true in our case, where we have about 20 cm/s or 50
peV spread in the vertical components of the arriving beam compared to a few
peV for the lowest vertically bound states inside the mirror-absorber
system~\cite{Westphal2} (see also~\cite{Prot}).

In addition, we have to introduce repopulation coefficients $p_j$ which allow
us to take into account a step between region IIa and IIb into account. If
there is no step, all $p_j$ =1. In the 1999 beam time, the $2^{nd}$ mirror has
been shifted downwards by 5 $\mu m$ relative to the $1^{st}$ mirror. The matrix
of overlap integrals of wave functions at the edge IIa/IIb is sufficiently
diagonal that we can neglect the off-diagonal elements. We can set $p_1 = 0.25$
and $p_j = 1$ for $j>1$. Hence, the relative shift of the bottom mirrors offers
a possibility of controlling the relative population of in particular the
ground state in the earth's gravitational field. However, even in the setup
without step, we find a reduction of the ground state for unknown reasons and
keep $p_1$  as a free parameter.

In the following, the roughness is described as an additional loss channel in
the one-dimensional Schr\"odinger equation. The neutron loss is then understood
in terms of non-specular scattering of neutrons into highly excited states
which due to their large vertical energy are rapidly lost inside the glass of
the mirror and the absorber/scatterer. Scattering happens at the rough and (due
to $V_{\rm absorber}\simeq 10^{-8}eV\gg E_{n}$ for states with low $n$) hard
surface of the absorber, which notion shall enable us to derive the
scattering-induced loss rate with just one undetermined micro-physical quantity
which would be given in a micro-physical scattering model as a function of the
scattering cross section. This parameter will be determined in the end by a fit
to the data.

The deformation of the wave functions compared with the purely gravitationally
bound states due to the large real part of the absorber potential leads to an
approximate vanishing of the bound states at the absorber surface. Therefore,
the loss rate is calculated in terms of the deformed states. In a second step,
one might model the neutron loss in terms of the physical parameters of the
absorber surface but this is outside of the scope of this article.

The neutron removal processes are modeled as a general phenomenological loss
rate $\Gamma_n(l)$ of the n$^{\rm th}$ bound state, which is taken to be
proportional to the probability density of the neutrons at the
absorber/scatterer. Here $l$ again denotes the position of the
absorber/scatterer above the mirror. The modulus of the bound state is then no
longer constant in time since it is given as the solution at first order of a
differential equation that determines the change of the norm to be proportional
to its momentarily value as well as the loss rate\beq
d\langle\psi_n|\psi_n\rangle=-\langle\psi_n|\psi_n\rangle\cdot\Gamma_n(l)\cdot
dt\;\;,\eeq which yields
\beq\langle\psi_n|\psi_n\rangle=|P_n(t)|^2\;,\;\;P_n(t)=e^{-\,\frac{1}{2}\,\Gamma_n(l)\cdot
t}\;\;.\label{flux0}\eeq

The roughness which causes the loss due to scattering can be thought of as
being confined to a region of about $2\cdot\sigma$ width attached to an
imagined absorber surface at a height $l$. Here $\sigma$ denotes the rms height
roughness of the absorber. Therefore, we give the loss rate of the n$^{\rm th}$
bound state in terms of the general description of scattering processes as a
function of the probability of neutrons to dwell within the roughness surface
region of the absorber/scatterer as
\begin{equation}
\Gamma_n(l)=\alpha_{\rm loss,n}\cdot\int_{l-2\sigma}^{l}dz |\psi_n(z)|^2
\;\;.\label{lossrate}
\end{equation}
Here we used the fact that the geometry of the wave guide system with and
without gravity allows for coordinates $(x,y,z)$ with $z$ denoting the
transverse coordinate for which the Schr\"odinger equation becomes separable
with the product ansatz
\begin{eqnarray}
\phi_n(\vec{r})&=&\psi_{n,g}(z)\cdot\phi_{xy}(x,y)\\
&&{\rm with: }\phi_{xy}(x,y)=\frac{1}{\sqrt{A_{xy}}}\cdot
e^{ik_xx+ik_yy}\;\;.\nonumber\label{wavefgen}
\end{eqnarray}
Since losses due to non-specular scattering should only depend on local
quantities of the surface and the probability of finding neutrons at the
surface, we assume that the micro-physics of neutron loss is independent of the
'macro'-physics of the wave function behavior. The above ansatz represents this
fact in its product structure which separates the micro-physical quantity
$\alpha_{\rm loss,n}$ completely from the behavior of the wave function
described by the probability density $|\psi_n(z)|^2$. Therefore, it is the
specific loss rate $\alpha_{\rm loss,n}$ which would depend in a micro-physical
model on the roughness properties $\sigma$ (roughness variance) and $\xi$
(roughness correlation length), i.e, $\alpha_{\rm loss,n}=\alpha_{\rm
loss,n}(\sigma,\xi)$. Furthermore this argument requires that micro-physical
quantities should not depend on the neutron state number. Thus, we have
$\alpha_{\rm loss,n}=\alpha_{\rm loss}\forall n$. Thus, we have
\begin{equation}
\Gamma_n(l)=\alpha_{\rm loss}\cdot\int_{l-2\sigma}^{l}dz
|\psi_n(z)|^2\;\;.\label{lossrate2}
\end{equation}
We fit this quantity $\alpha_{\rm loss}$ to the data.

This leaves us with the task to determine the integrals $\int_{l-2\sigma}^{l}dz
|\psi_n(z)|^2$ for a state $n$ in a given experimental setup.

\subsection{The wave guide system with
gravity}\label{grav}

The bound states  $\psi_{n,g}$ in the linear gravitational potential are
confined by two very high potential steps above (absorber) and below (mirror).
They can be given analytically as:
\begin{equation}
\psi_{n,g}(z)=A_n\cdot Ai\left(z/R-\zeta_n(l)\right)+B_n\cdot
Bi\left(z/R-\zeta_n(l)\right) \label{Eq.30}
\end{equation}
and they fulfill the boundary conditions:
\begin{equation}
\psi_{n,g}\Big|_{z=0}=0\;\;\wedge\;\;\psi_{n,g}\Big|_{z=l}=0 \label{Eq.31}
\end{equation}
which account for the high potential steps bounding the potential from below
and above. Eq.s~(\ref{Eq.30}) and (\ref{Eq.31}) together determine the energy
eigenvalues $\zeta_n(l)$ as functions of the absorber height $l$ which is done
numerically.

For a given state $\psi_{n,g}$ the $Ai$-part in the wave function is
exponentially decaying for $z>z_n(l)=\zeta_n(l) R$ while the $Bi$-part grows
exponentially in this region. At $z=l$ both parts add up to zero. As long as
$l>z_n(l)$ we have $B_n<<A_n$ because the $Bi$-part has to compensate only for
the exponentially small value of the $Ai$-part at $z=l$. This, however, implies
that at $z<l$ we have $|B_n Bi(z/R-\zeta_n(l))|<A_n Ai(z/R-\zeta_n(l))$.
Therefore, we can ignore the $Bi$-part of the wave-function in the calculation
of an asymptotic expression for the $\Gamma_{n,g}(l)$ at large $l$. As a
result, we have \beq\psi_{n,g}(z)\approx A_n \cdot
Ai\left(z/R-\zeta_n(l)\right)\label{Eq.30b}\eeq at $l\gg z_n$. The main effect
of the $Bi$-part is the increase of $\zeta_n(l)>z_n$ at $l<z_n$ where the $z_n$
are given by Eq.~\ref{Eq.06b}. Here we denote all quantities at $l\to\infty$
with $\zeta_n,A_n$ etc. while the corresponding quantities of the realistic
states Eq.~\ref{Eq.30} at finite $l$ are denoted with $\zeta_n(l),A_n(l)$ etc.

The states $\psi_{n,g}$ are well approximated in the large $l$ regime by the
asymptotic WKB states
\begin{eqnarray}
\psi_{n,g}\approx\frac{A_n}{2\sqrt{\pi
R}}\cdot\left(\zeta-\zeta_n\right)^{-1/4}\cdot e^{-\frac{2}{3}\left(\zeta-\zeta_n\right)^{3/2}}\nonumber\\
\label{Ailargein}\\
\mathrm{if}\quad \zeta>\zeta_n\;.\nonumber
\end{eqnarray}
In the limit $l\to\infty$ we have the $\zeta_n(l)\to\zeta_n$ given by
Eq.~\ref{Eq.06b} to $1^{\rm st}$ order in WKB and $A_n=\pi Bi(-\zeta_n)$. (For
$l\to\infty$ we have $\psi_{n,g}(z)=A_n\cdot Ai(z/R-\zeta_n)$. Then the quoted
value of $A_n$ which is exact ensures that
$\langle\psi_{n,g}|\psi_{n,g}\rangle=1$.) In the case $l=R\zeta_l\lesim z_n$
consider the WKB-expression for the energy eigenvalues \beq
\int_0^{\zeta_l<\zeta_n(l)} d\zeta\sqrt{\zeta_n(l)-\zeta}=\pi (n-1/4)\;\;.\eeq
With \beq \int_0^{\zeta_l<\zeta_n(l)}
d\zeta\sqrt{\zeta_n(l)-\zeta}=\zeta_l\sqrt{\zeta_n(l)}+{\cal
O}\left(\frac{\zeta_l^2}{\zeta_n^2(l)}\right)\eeq we arrive at \beq
\zeta_n(l)=\frac{\pi^2 (n-1/4)^2}{\zeta_l^2}\eeq which coincides with the known
box state expression for large $n$.

Now plugging the asymptotics of the gravitationally bound states
Eq.~\ref{Ailargein} into Eq.~\ref{lossrate2} one arrives at a prediction for
the loss rate that reads
\begin{eqnarray}
\Gamma_{n,g}(l)&=&\alpha_{\rm
loss}\cdot\int_{l-2\sigma}^{l}dz\cdot|\psi_{n,g}(z)|^2\nonumber\\
&\simeq&\alpha_{\rm loss}\cdot\frac{\sigma}{2\pi R}\cdot
|A_n|^2\cdot
\frac{e^{-\,\frac{4}{3}\left(\frac{l-z_n}{R}\right)^{3/2}}}{\sqrt{(l-z_n)/R}}
\hspace{5.0ex}{\rm for}\;\;\;l\gg z_n\;.\label{loss1}
\end{eqnarray}

Now look at the behavior of the total neutron flux $F$ through the wave guide
if a large number of states contribute to it (semi-classical limit). We have
$F=\int_{\cal A}j_x\sim\sum_n\langle\psi_n|\psi_n\rangle$ where we get
$\langle\psi_n|\psi_n\rangle$ from Eq.~\ref{flux0} and ${\cal A}$ denotes the
wave guide cross section. If gravity is present the behavior $\Gamma_{\rm
loss}\sim \exp(-4/3\cdot(\zeta-\zeta_n(l))^{3/2})$ leads to the fact that each
time when $l\simeq z_n=R\zeta_n$ a new state rapidly starts to contribute to
the transmission. Therefore at a given large height $l$ the number of states
contributing to $\phi(l)$ reads from Eq.~\ref{Eq.06b}
\[l=z_n\sim N^{2/3}\Rightarrow N(l)\sim l^{3/2}\] which yields
asymptotically (if $N(l)$ large and thus $\Delta N_{N-(N-1)}/N= 1/N\rightarrow
0$) the classical behavior in a gravitational field. A simple phase space
argument \cite{Nesvizhevsky2} shows that a perfect absorber at the top in
presence of a gravitational field yields a classical transmission\[F(l)\sim
l^{3/2}\;\;.\]

\subsection{A wave-guide system without gravity}

Now we can use the model to derive the loss rate for bound states in the
absence of gravity. These bound states are well approximated by those which
describe the quantum dynamics of a particle in a one-dimensional box with
infinitely high walls, i.e. the so-called box states. They are given by
\begin{eqnarray}
\psi_n&=&\sqrt{\frac{2}{l}}\cdot \sin(\frac{n\pi}{l}\cdot z)\label{curr4}
\end{eqnarray}
which yields a loss rate given by
\begin{eqnarray}
\Gamma_n(l)&=&\alpha_{\rm
loss}\cdot\int_{l-2\sigma}^{l}dz\cdot|\psi_n(z)|^2\nonumber\\
&=&\alpha_{\rm
loss}\cdot\left[\frac{2\sigma}{l}-\frac{1}{2n\pi}\sin\left(
\frac{4n\pi}{l}\cdot\sigma\right)\right]\;\;.\label{loss2}
\end{eqnarray}
A comparison of the full box state expression (line 2 in the above
equation) with the numerical result for the loss rate with gravity
using the full states Eq.~\ref{Eq.30} is given graphically in
Fig.~\ref{Fig.3}.

The exact expression in the second line of Eq.~\ref{loss2}
approaches a constant for $n\to\infty$ rendering the sum in
Eq.~\ref{boxfluxasympt} divergent. Therefore, in a realistic fit
we have to include the fact that in any real experiment the number
of box states $N$ in the wave guide is finite. Firstly, the
collimator system in front of the wave guide yields an input
vertical velocity distribution of finite width. Secondly, for
wider vertical velocity spectra all neutrons with $v_z>v_z^{\rm
crit}=4.3\cdot m\cdot s^{-1}$, the critical velocity of glass,
will enter the mirror or the absorber directly without forming
bound states in the wave guide. In both cases the number of box
states which is populated by the entering flux of neutrons behaves
like $N\sim l$. If the collimator is tuned to yield an input
vertical velocity distribution of small width (i.e. about
$10\,cm/s$) we have $N/l\sim 2\cdot\mu m^{-1}$ corresponding to
about $N\approx 200$ box states in a wave guide of $l\sim 100\,\mu
m$ width. Thus, we have evaluated a finite sum with $N/l=2\cdot\mu
m^{-1}$ when comparing the gravity-free prediction of
Eq.~\ref{boxfluxasympt} with the experimental data sets (for
$l<100\,\mu m$ we have $N<200$ box states populated with $N$
approaching $\approx 200$ for $l\to 100\,\mu m$). The dependence
of our result on the choice of the cutoff $N$ is very weak. If the
critical velocity of glass defines the cutoff this results in
$N/l\sim 20\cdot\mu m^{-1}$ (corresponding to about $N\approx
2\cdot 10^3$ box states in a wave guide of $l\sim 100\,\mu m$
width). We find that the predictions agree for both cutoff choices
with each other to far better than the experimental accuracy
within the $l$-range of the measurement.

Now in the absence of gravity the asymptotic behavior of the
transmitted flux $F(l)$ carried by the box states at large $l$ can
be given directly from the Eq.s~\ref{flux0} and \ref{loss2}. For
$n\sigma/l\ll 1$ we can approximate Eq.~\ref{loss2} with \beq
\Gamma_n(l)=\alpha_{\rm loss}\cdot16\pi^2/3\cdot
\sigma^3\cdot\frac{n^2}{l^3}\quad\quad{\rm if:}\quad\frac{n
\sigma}{l}\ll 1\;\;. \label{loss2approx}\eeq If the approximation
in Eq.~\ref{loss2approx} were valid for all $n$ we would have
\begin{eqnarray}
F(l)&\sim&\sum_{n\geq1}e^{-\Gamma_n(l)\cdot t_{\rm
flight}}\nonumber\\
&=&\sum_{n\geq1}e^{-\gamma\cdot n^2}\;,\;\;{\rm
with:}\;\gamma=16\pi^2/3\cdot \sigma^3\cdot l^{-3}\cdot t_{\rm
flight}\nonumber\\
&=&\frac{1}{2}\cdot\left(\vartheta_3(0,e^{-\gamma})-1\right)\nonumber\\
&\sim&l^{3/2}\quad\quad{\rm for}\;{\rm large}\;l\;\;.\label{boxfluxasympt}
\end{eqnarray}
Here $\vartheta_n(q,u)$ denotes the elliptic theta function where
we used {\it Mathematica}~\cite{wol} to evaluate the sum. This
result differs from the naive classical behavior of a gravity-free
wave guide with a perfect absorber: In the case of the linear
trajectories describing classical particles in absence of the
gravitational field, we find \cite{Westphal2} \beq F(l)\sim
l^2\eeq which is easy to imagine since one factor of $l$ obviously
has its origin in the relation \hbox{$F(l)\sim {\cal A}\sim l$}
while the $2^{\rm nd}$ factor encodes that the range of vertical
velocities $\Delta v_z$ of particles which pass the wave guide
without ever touching the absorber also behaves like $\Delta
v_z\sim l$.

Finally, from this situation we expect in general an interpolating
behavior of the gravity-free transmission rate with respect to its
power-law dependence on the absorber height. In fact, if the
gravity-free prediction of Eq.~\ref{boxfluxasympt} is carried out
using the exact expression for the loss rates $\Gamma_n$ in
Eq.~\ref{loss2} for $N/l>200\cdot\mu m^{-1}$, i.e. $N>2\cdot 10^4$
box states at $l\approx 100\,\mu m$, we find that the transmission
begins to deviate from an $l^{3/2}$-power law towards an
$l^n$-dependence with $n\to 2$ which is the general dependence to
be expected both classically and quantum mechanically. Thus, the
behavior of the gravity-free prediction as $F(l)\sim l^{3/2}$ in
our given experimental situation is an artifact caused by the
relatively small number of box states ($N\lesim 2\cdot 10^3$ for
$l\leq 100\cdot\mu m$) in the wave guide.

\subsection{Reversed geometry}\label{reverse}

We now turn to the third case of $g\to -g$ instead of $g\to 0$. This inversion
of gravity is equivalent to a setup geometry where the absorber/scatterer is
placed at the bottom at $z=0$ and a movable mirror at $z=l$ above the absorber.
For this situation we can follow the derivation of Subsection~\ref{grav}. Since
the absorber is now at $z=0$ we have to evaluate the probability integral at
this position which implies for large $l$ the use of the asymptotic WKB
expression
\begin{eqnarray}
\psi_{n,g}\approx\frac{A_n}{\sqrt{\pi
R}}\cdot\left(\zeta_n-\zeta\right)^{-1/4}\cdot\sin\left[\frac{2}{3}\left(\zeta_n-\zeta\right)^{3/2}+\frac{\pi}{4}\right]\nonumber\\
\label{Aismallin}\\
\mathrm{if}\quad \zeta<\zeta_n\;.\nonumber
\end{eqnarray}
This results in
\begin{equation}
\Gamma_{\rm loss}^{(n,g,\rm rev.)}(l)=\alpha_{\rm loss}\cdot\frac{\sigma}{2\pi
R}\cdot
|A_n|^2\cdot\frac{16}{3}\sqrt{\frac{z_n}{R}}\cdot\frac{\sigma^2}{R^2}\;\;.\label{loss3}
\end{equation}
Note, that this loss rate of the reversed geometry is practically
independent on the state number $n$ since
$\lim_{n\to\infty}|A_n|^2\sqrt{\zeta_n}=\pi$ and even at $n=1$ it
is $|A_1|^2\sqrt{\zeta_1}/\pi-1\lesim 0.5\%$. Comparing this
result with the corresponding expression Eq.~\ref{loss1} for the
normal geometry we find that the ratio of the fluxes $F_n^{(g),\rm
rev.}(l)/F_n^{(g)}(l)$ of the $n^{\rm th}$ state for $l>z_n$
between the reversed and the normal geometry is given by \beq
\frac{F_n^{(g),\rm rev.}(l)}{F_n^{(g)}(l)}=e^{-\alpha_{\rm
loss}\cdot\frac{\sigma}{2\pi R}\cdot
|A_n|^2\cdot\frac{16}{3}\sqrt{\frac{z_n}{R}}\cdot\frac{\sigma^2}{R^2}
\cdot\frac{L}{v_{hor.}}}\label{ratio}\;\;\;\;.\eeq Since
$\frac{\sigma}{2\pi R}\cdot
|A_1|^2\cdot\frac{16}{3}\sqrt{\frac{z_1}{R}}\cdot\frac{\sigma^2}{R^2}
\cdot\frac{L}{v_{hor.}}\approx 7.2\cdot 10^{-5}s$ for experimental
values of $L=0.13\cdot m$, $v_{hor.}\approx 10\cdot m/s$ and
$\sigma=0.75\,\mu m$ a value of $\alpha_{\rm loss}\gesim{\cal
O}(10^4 s^{-1})$ fitted from a measurement of the normal geometry
would result in a huge asymmetry under a $\pi$-rotation around the
optical axis of the wave guide when comparing the normal and the
reversed geometry.

\section{A fit to the data}

For a comparison with the measurements we plug the resulting loss rates into
the general prediction for the transmitted neutron flux Eq.~\ref{flux}.
Together with the repopulation coefficient $p_1$ accounting for eventual shifts
between split bottom mirrors one predicts

\begin{eqnarray}
F(l)&=&F_0+\sum_n F_n(l)=F_0+C\cdot\left\{p_1\cdot e^{-\Gamma_{\rm
loss}^{(1)}(l)\cdot\frac{L}{v_{hor.}}}\space+\sum_{n>1}\cdot e^{-\Gamma_{\rm
loss}^{(n)}(l)\cdot\frac{L}{v_{hor.}}}\right\}\label{flux}
\end{eqnarray}
$F_0$ is the detector background and $C$ the total flux normalization.
$\alpha_{\rm loss}$ is the universal parameter introduced above which
parameterizes the scatterer strength. This quantity is in general expected to
be a rather weak function of the roughness parameters $\sigma$ and $\xi$ which
would be determined in principle by a microphysical calculation of the loss due
to non-specular scattering at the rough absorber surface. Thus it does not
depend on the absence or presence of gravity nor on the state number $n$.

In the case of $l\lesim z_n$, the absorber/scatterer will begin to squeeze the
bound states once they start to 'feel' it sufficiently strongly. This implies
further that due to $E_{n,g}(l)\geq E_{n,g}(\infty)=E_{n,g\,{\rm pure\;
gravity}}$ for the true bound states a sufficiently small $l$ leads to
$E_{n,g}(l)\gg mg\cdot l$. Therefore, the calculation of the $\Gamma_{\rm
loss}^{(n)}$ is done using the full realistic bound states Eq.~\ref{Eq.30} and
deriving the relations corresponding to Eq.s~\ref{loss1}, \ref{loss2} and
\ref{loss3} numerically which incorporates the mentioned behavior.

\subsection{Normal geometry}

The detector background $\phi_0$ has been measured independently to yield
$F_0=(0.0043\pm0.0004)s^{-1}$ for the 1999 measurement \cite{Nesvizhevsky1} and
$F_0=(0.0004\pm0.0001)s^{-1}$ for the new 2002 measurement using an improved
setup \cite{Nes05}. Thus, one remains with having to determine the two
universal quantities $\alpha_{\rm loss}$ and $C$ from the data. $C$ turns out
to be completely fixed by the data points at $l>70\,\mu m$, and thus also has
been measured. The fit therefore will be a 1-parameter one determining
$\alpha_{\rm loss}$ as long as all the populations  $p_n$ stay to be equal.

We fit now Eq.~\ref{flux} to the newer data from the run of the
experiment in 2002 which has a different absorber and better
statistics, and systematic effects are smaller than in 1999. The
results are shown in Fig. 2. The value of $\alpha_{\rm loss}$ is
found in a fit to the data. The result of the fit yields
($L=13\,{\rm cm}$, $v_{hor.}\approx 5\cdot m\cdot s^{-1}$)
\begin{equation}
\alpha_{\rm loss}=\,(3.4\pm0.1)\cdot10^4 s^{-1}\;\;.\label{alphfit}
\end{equation}
The fit was done using neutrons with only one value of the
horizontal velocity which was chosen to be the average velocity
$v_{hor.}\approx 5\cdot m\cdot s^{-1}$. This approximation
produces essentially the same results as if one uses the full
actual spectrum of horizontal velocities produced by the
collimating system to calculate the transmission.

The 2002 run was performed with only one bottom mirror so that no
significant repopulation effects of the ground state are expected.
However, it was necessary to allow the population of the ground
state to shift towards 75\% compared to the excited states in
order to describe the data: $p_1=0.77\pm 0.04$. This suppression
is significant, but it is not large leaving all the states still
to be approximately equally populated.

In the measurement of the 1999 run, two bottom mirrors were used.
It was tried to shift these mirrors relative to each other
vertically and in alignment by a few micrometers. Small shifts of
a few microns between two bottom mirrors change the population of
the ground state and the next state quite drastically: A shifted
geometry with a relative mirror shift of about $5\;\mu m$ and no
relative tilt of the two bottom mirrors results in $p_1\approx
0.25$ compared to $p_n=1$, $n\geq 2$. The fit to the data (see
Fig.~\ref{Fig.4}) results in $\alpha_{\rm
loss}=\,(5.3\pm0.5)\cdot10^4 s^{-1}$ and $p_1=0.24\pm0.1$ which
would be consistent with the possible relative mirror shift
discussed above.

\subsection{Reversed geometry}

\begin{figure}[ht]
\begin{minipage}{18pc}
\includegraphics[width=18pc]{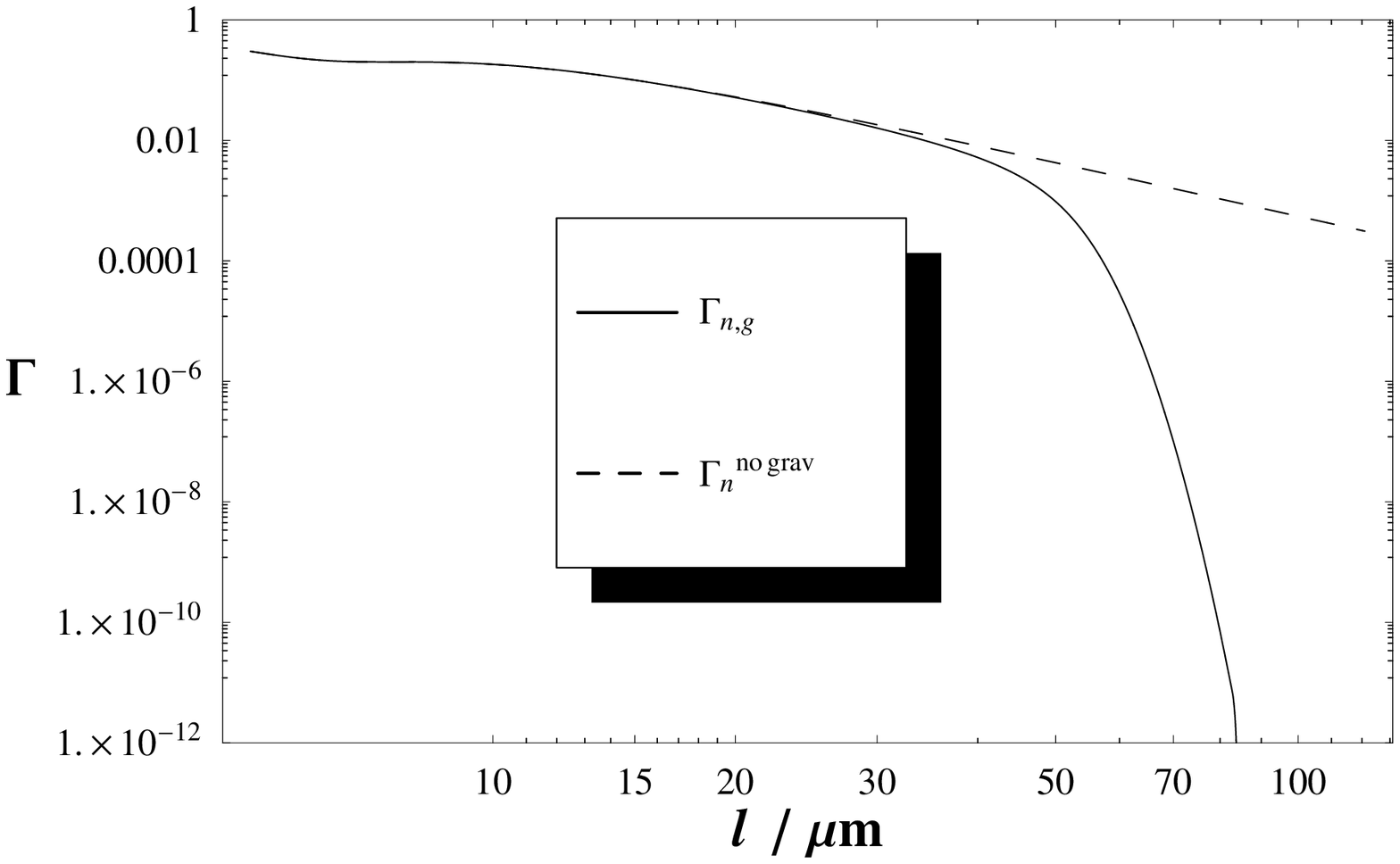}
\caption{\label{Fig.3}\small Behavior of the loss rate $\Gamma$
(plotted in arbitrary units) for the pure box state $n=5$ (dash),
an approximate power law, and for the gravitationally bound state
$n=5$ found numerically from using the full states Eq.~\ref{Eq.30}
(solid), showing exponentially fast decay above some $l={\cal
O}(R)$. }
\end{minipage}\hspace{1pc}
\begin{minipage}{18pc}
\includegraphics[width=18pc]{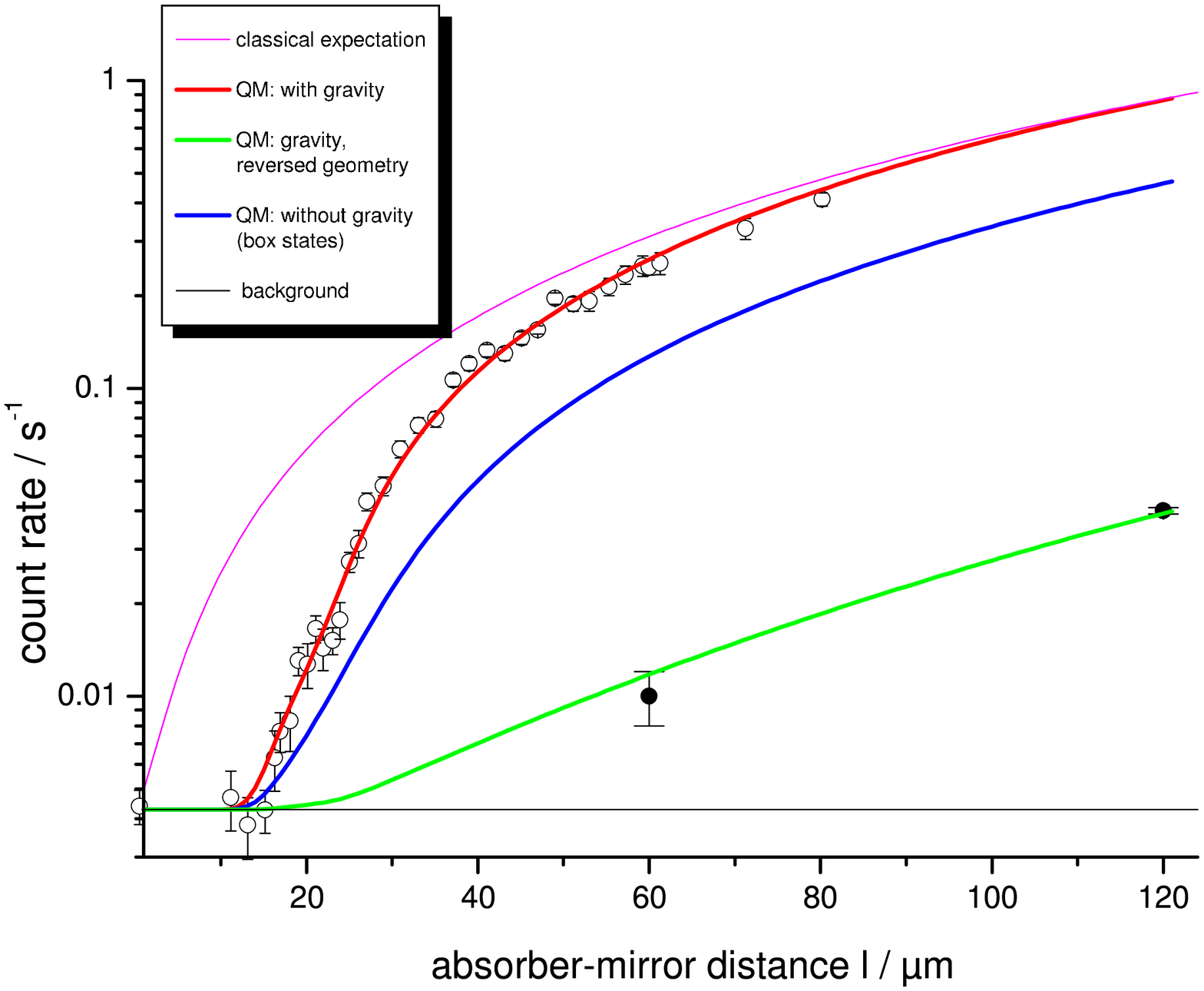}
\caption{\label{Fig.4}\small Open Circles: Transmission as a
function of the absorber-mirror distance. Filled Circles:
Transmission measured with inverse geometry - the absorber on the
bottom and a 10 cm mirror on top (both are 1999 measurement
data~\cite{Nesvizhevsky2}).}
\end{minipage}
\end{figure}

In a second setup described by the term 'reversed geometry', the
absorber was placed at the bottom and the mirror above. Here, the
position where the scattering-inducing roughness is found at
$z=0$. This case can also be derived from the states
Eq.~\ref{Eq.30} by just placing the roughness appropriately on
bottom and then calculating the corresponding loss rates. The
bound states are now again given by the Airy functions. However,
at $z=0$ where the absorber is now placed, they do not decay
exponentially fast and the gravitationally bound states will be
strongly absorbed at arbitrary heights $l$ of the mirror at top -
quite contrary to the normal setup. If gravity were absent, such a
$\pi$-turn of the wave guide around its optical axis would  not
have such an effect. Thus, there would be no preferred direction
in space forcing the transmission factor to be invariant under
rotations around the optical axis of the system since the box
states describing the simple confinement situation without gravity
are symmetric with respect to the optical axis. This check was
done with the 1999 data. We took the prediction for
gravitationally bound states of Eq.~\ref{Eq.30} in the normal
setup and fitted $\alpha_{\rm loss}$ and $p_1$ to the data. The
model yields $p_1=0.24\pm 0.1$ and $\alpha_{\rm
loss}=\,(5.3\pm0.5)\cdot10^4 s^{-1}$. Taking now these values for
$\alpha_{\rm loss}$ and $p_1$ from the fit, we can calculate a
prediction for the setup with reversed geometry using the results
of Subsection~\ref{reverse} - which is now entirely fixed and not
fitted any more. Using the experimental setup parameters used
there, $L=13\,cm$ and $v_{\rm hor.}\approx 10\cdot m/s$ we can
estimate the suppression factor using Eq.~\ref{ratio} on the
ground state which yields \beq \frac{F_1^{(g),\rm
rev.}(l)}{F_1^{(g)}(l)}\approx 0.03\;\;\;\;.\eeq This fits well
with the two data points obtained experimentally in this reversed
setup. Using the full prediction calculated again numerically from
Eq.~\ref{Eq.30} in the reversed setup, the comparison with the
1999 data is shown in Fig.~\ref{Fig.4}. In treating the 1999 data
other models~\cite{Abele,Nesvizhevsky2} showed more pronounced
steps. These steps are not present in the (more advanced) model
presented in this paper.

The above described asymmetry of the neutron transmission of a
wave guide with one absorbing and one reflecting wall under
rotations around its optical axis in the presence of gravity has
indeed been measured and can be described quantitatively. This
rules out the possibility to explain the data just by confinement
effects in terms of box states inside a rectangular box-shaped
potential \cite{Hansson} and thus establishes the gravitational
nature of force which binds the neutron bound states to the bottom
mirror. For further discussion, see also \cite{Nes03}. The
measurement of the reverse geometry is a very good test of our (or
probably any) model, since the result depends strongly on the
absorber properties. The fact that the absorber model which
describes the data in the normal geometry can also be used for the
reversed geometry, without any new fit parameters, gives us
confidence that our model is a useful tool to describe the data.

\section{Summary}

For the first time the existence of quantized bound states of neutrons in the
gravitational field of the earth above a horizontal glass mirror was
experimentally demonstrated. Here we present a quantum mechanical model
providing an accurate fit to the data. The difficult part is the incorporation
of the neutron absorber into this model. We show that we can describe its
action with only one phenomenological fit parameter in several configurations.
The most important configuration has bound quantum states in the linear
gravitational potential. We consider repopulation effects of these states when
a step from a second mirror is present. The emerging reduction of the first
state has been calculated. The standard setup uses only one mirror. Also in
this case, the data show an unexpected reduction of the first quantum state as
well.

Further, it is shown that the experiment would generate different results, if
gravity were hypothetically to be turned off. This configuration is a
wave-guide system with box states which describe the quantized motion of a
particle a one-dimensional box with infinitely high walls.

A striking difference is found, if the gravitational field changes sign. This
last case has been observed by our measurement~\cite{Nesvizhevsky1} and can be
described quantitatively within the same modeling of the absorber mechanism
presented here. Thus, the dominating effect of gravity on the formation of the
bound states has been demonstrated since the measurement did not show just a
simple confinement effect, as proposed in~\cite{Hansson}. On the contrary, it
could not be described theoretically, if the earth's gravitational field was
not present.

% Produces the bibliography via
%\item name(s) and initials;
%\item date published;
%\item title of journal, book or other publication;
%\item titles of journal articles may also be included (optional);
%\item volume number;
%\item editors, if any;
%\item town of publication and publisher in parentheses for {\it books};
%\item the page numbers.
%\end{itemize}

\end{document}